\documentclass[%
 reprint,
 amsmath,amssymb,
 aps,
]{revtex4-1}

\usepackage{graphicx}% Include figure files
\usepackage{dcolumn}% Align table columns on decimal point
\usepackage{bm}% bold math
\usepackage[linktocpage=true,
  colorlinks=true, 
  pdfborder={0 0 0},
  linkcolor=blue,
  citecolor=red,
  filecolor=yellow,
  urlcolor=blue,
  bookmarks,
  pdfauthor={},
]{hyperref}

\begin{document}

\preprint{APS/123-QED}

\title[Wannier interpolation of GF]{Wannier interpolation of one-particle Green's functions from\\
coupled-cluster singles and doubles (CCSD)}

\author{Taichi Kosugi}
\email{kosugi.taichi@gmail.com}
\affiliation{
Department of Physics,
University of Tokyo,
Tokyo 113-0033,
Japan 
}

\author{Yu-ichiro Matsushita}
\affiliation{
Laboratory for Materials and Structures,
Institute of Innovative Research,
Tokyo Institute of Technology,
Yokohama 226-8503,
Japan
}

\date{\today}

\begin{abstract}
We propose two schemes for interpolation of the one-particle Green's function (GF) calculated within coupled-cluster singles and doubles (CCSD) method for a periodic system.
They use Wannier orbitals for circumventing huge cost for a large number of sampled $k$ points.
One of the schemes is the direct interpolation,
which obtains the GF straightforwardly by using Fourier transformation.
The other is the self-energy-mediated interpolation,
which obtains the GF via the Dyson equation.
We apply the schemes to a LiH chain and $trans$-polyacetylene
and examine their validity in detail.
It is demonstrated that the direct-interpolated GFs suffer from numerical artifacts stemming from slow convergence of CCSD GFs in real space,
while the self-energy-mediated interpolation provides more physically appropriate GFs due to the localized nature of CCSD self-energies.
Our schemes are also applicable to other correlated methods capable of providing GFs.
\end{abstract}

\maketitle

%\tableofcontents

\section{introduction}
\label{sec:intro}

Although electronic-structure calculations based on the density functional theory (DFT)\cite{bib:76,bib:77} have been successful by and large for quantitative explanations and predictions of the properties of molecules and solids,
they are known to have a tendency to fail in describing the material properties even qualitatively for strongly correlated systems.
To remedy such shortcomings of DFT,
various approaches have been proposed.
There exist such approaches based on the Green's function (GF) theory,
including $GW$ method.\cite{bib:GW1,bib:GW2,bib:GW3}
They often use the non-interacting states obtained in DFT calculations as reference states for the construction of interacting GFs.
On the other hand,
many sophisticated approaches based on the wave function theory have been developed for quantum chemistry calculations.
The coupled-cluster singles and doubles (CCSD) method\cite{Helgaker} is a widely accepted one
since it achieves moderate balance between its high accuracy and high computational cost.
Not only is the relation between $GW$ and CCSD methods theoretically interesting,
but also their quantitative comparison is worth examining\cite{bib:4535} from a practical viewpoint.

Photoelectron spectroscopy is one of the most active fields in experimental physics of today.
Measurements of the photoelectric effects in target materials make use of various kinds of techniques such as angle-resolved photoemission spectroscopy (ARPES) for clarifying the material properties.
The measured spectra of an interacting electronic system are often explained under a certain assumption via the one-particle GF.\cite{bib:4070,bib:4165,bib:pw_unfolding}
The clear understanding of the characteristics of GFs is thus important both for theoretical and practical studies in material science.
Mathematically speaking,
the quasiparticle and satellite peaks in photoelectron spectra represent nothing but the poles of one-particle GF of an interacting system.
Particularly, the distance between the peaks closest to zero frequency is the fundamental gap.
It has been demonstrated that there exists an analytically solvable model\cite{bib:4575} which helps to obtain transparent insights into interacting GFs.
Meanwhile, the GFs in the context of correlated electronic-structure calculations for
uniform electron gases\cite{bib:4033} and realistic systems have been drawing attention recently\cite{bib:4473,bib:4483,bib:4516,bib:4582},
which we deal with in the present study.

CCSD\cite{bib:4115} and subsequent GF calculations\cite{Nooijen92,Nooijen93,Nooijen95,bib:3947,bib:4275} are difficult especially for a periodic system due to their large computational cost since a sufficiently large number of sampled $k$ points is needed.
This fact hinders one from performing detailed comparison between the band structures obtained by a Hartree--Fock (HF) or DFT calculation and
the spectra obtained from CCSD GF,
and the measured spectra.
Development of physically appropriate interpolation schemes for CCSD GFs is thus desirable for examining spectral properties of correlated systems,
which is nothing but what we do in this study.

This paper is organized as follows.
In Sect. \ref{sec:method},
we review CCSD and GF calculations briefly and
explain the interpolation schemes.
In Sect. \ref{sec:comp_details},
we describe the details of our computation.
In Sect. \ref{sec:results},
we show the results for the target systems.
In Sect. \ref{sec:conclusions},
our conclusions are provided.

\section{method}
\label{sec:method}

\subsection{CCSD and GF for a periodic system}

The CC state for a reference state $| \Psi_0 \rangle$ is constructed by performing an exponentially parametrized transform as $| \Psi_{\mathrm{CC}} \rangle \ = e^{\hat{T}} | \Psi_0 \rangle$,
where $\hat{T}$ is a so-called cluster operator.
The normalization of our CCSD wave functions obeys the bi-variational formulation,\cite{Arponen83,Bi-vari1,Bi-vari2}
with which we calculate the CCSD one-particle GFs\cite{Nooijen92,Nooijen93,Nooijen95} in the recently proposed procedure\cite{bib:3947,bib:4275} as well as in our previous studies.\cite{bib:4473,bib:4483,bib:4516}

Here we review briefly the calculation of CCSD GF for a periodic system.
The GF in frequency domain is given by
\begin{gather}
    G (\boldsymbol{k}, \omega)
    =
        G^{(\mathrm{h})} (\boldsymbol{k}, \omega)
        +
        G^{(\mathrm{e})} (\boldsymbol{k}, \omega)
    ,
    \label{def_CCSD_GF}
\end{gather}
where
\begin{gather}
    G_{p p'}^{(\mathrm{h})} (\boldsymbol{k}, \omega)
    =
        \langle \Psi_0 |
        (1 + \hat{\Lambda})
        \overline{a_{\boldsymbol{k} p}^\dagger}
        \frac{1}{\omega + \overline{H}}
        \overline{a}_{\boldsymbol{k} p'}
        | \Psi_0 \rangle
    \label{def_G_hole}
\end{gather}
and
\begin{gather}
    G_{p p'}^{(\mathrm{e})} (\boldsymbol{k}, \omega)
    =
        \langle \Psi_0 |
        (1 + \hat{\Lambda})
        \overline{a}_{\boldsymbol{k} p}
        \frac{1}{\omega - \overline{H}}
        \overline{a_{\boldsymbol{k} p'}^\dagger}
        | \Psi_0 \rangle
    \label{def_G_elec}
\end{gather}
are the partial GFs from the hole and electron excitations,
respectively.
$\boldsymbol{k}$ is a wave vector and $\omega$ is a complex frequency.
$p$ is the composite index of a spatial orbital and a spin direction for an occupied or unoccupied single-electron state. 
For the original Hamiltonian $\hat{H}$,
we defined the similarity transformed Hamiltonian 
$
\overline{H}
\equiv
e^{-\hat{T}} \hat{H} e^{\hat{T}} - E_0
$
measured from the CCSD total energy $E_0$.
We also defined the transformed creation and annihilation operators
$
\overline{a_{\boldsymbol{k} p}^\dagger}
=
e^{-\hat{T}}
\hat{a}_{\boldsymbol{k} p}^{\dagger}
e^{\hat{T}}
$
and
$
\bar{a}_{\boldsymbol{k} p}
=
e^{-\hat{T}}
\hat{a}_{\boldsymbol{k} p}
e^{\hat{T}}
,
$
respectively.
$\hat{\Lambda}$ is the parametrized de-excitation operator determined in the $\Lambda$-CCSD calculation,\cite{bib:3947,bib:4275}
which has to be introduced since the CCSD operator $e^{\hat{T}}$ is not unitary.

In order to avoid the computational difficulty in treating the inverse matrix $(\omega \pm \overline{H})^{-1}$
in eqs.  (\ref{def_G_hole}) and (\ref{def_G_elec}),
the parametrized operators
$\hat{X}_{\boldsymbol{k} p} (\omega)$
and
$\hat{Y}_{\boldsymbol{k} p} (\omega)$
are introduced so that\cite{bib:3947,bib:4275}
\begin{gather}
    (\omega + \overline{H})
    \hat{X}_{\boldsymbol{k} p} (\omega)
    | \Psi_0 \rangle
    =
        \overline{a}_{\boldsymbol{k} p}
        | \Psi_0 \rangle
    \label{def_ip_eom}
\end{gather}
and
\begin{gather}
    (\omega - \overline{H})
    \hat{Y}_{\boldsymbol{k} p} (\omega)
    | \Psi_0 \rangle
    =
        \overline{a_{\boldsymbol{k} p}^\dagger}
        | \Psi_0 \rangle
    .
    \label{def_ea_eom}
\end{gather}
The linear equation for the non-Hermitian matrix in eq. (\ref{def_ip_eom}) is called the ionization potential (IP) equation-of-motion (EOM) CCSD equation,
while that in eq. (\ref{def_ea_eom}) is called the electron affinity (EA) EOM-CCSD equation.
After obtaining the parametrized operators,
we use them in eqs (\ref{def_G_hole}) and (\ref{def_G_elec}) to get
\begin{gather}
    G_{p p'}^{(\mathrm{h})} (\boldsymbol{k}, \omega)
    =
        \langle \Psi_0 |
        (1 + \hat{\Lambda})
        \overline{a_{\boldsymbol{k} p}^\dagger}
        \hat{X}_{\boldsymbol{k} p'} (\omega)
        | \Psi_0 \rangle
    \label{G_hole_X}
\end{gather}
and
\begin{gather}
    G_{p p'}^{(\mathrm{e})} (\boldsymbol{k}, \omega)
    =
        \langle \Psi_0 |
        (1 + \hat{\Lambda})
        \overline{a}_{\boldsymbol{k} p}
        \hat{Y}_{\boldsymbol{k} p'} (\omega)
        | \Psi_0 \rangle
    .
    \label{G_elec_Y}
\end{gather}

The $k$-resolved spectral function is defined via the GF as
\begin{gather}
    A (\boldsymbol{k}, \omega)
    =
        -\frac{1}{\pi}
        \mathrm{Im Tr} \,
        G (\boldsymbol{k}, \omega + i \delta)
    \label{def_spec_k}        
\end{gather}
for a real $\omega$ with a small positive constant $\delta$ ensuring causality.
The spectral function calculated in this way reflects our correlated approach,
to be compared with the band structures obtained in mean-field-like approaches such as HF and DFT.

Before moving on to the description of our interpolation schemes,
it is noted here that
there exists an alternative to obtain correlated spectra or band structure for arbitrary $k$ points without resorting to interpolation.
Specifically, 
usage of a large series of shifted regular $k$ meshes
enables one to perform EOM-CCSD calculations to get the excitation energies for an arbitrarily fine $k$ mesh,
as adopted by McClain et al.\cite{bib:4115}
This approach requires large computational cost for the accuracy ensured by the EOM-CCSD framework itself.

\subsection{Wannier interpolation}

\subsubsection{Wannier orbitals}

Wannier orbitals (WOs)\cite{bib:5} and their variants in solids are analogues of Foster--Boys orbitals\cite{bib:4594,bib:4595} in molecular systems.
In particular, maximally localized WOs (MLWOs)\cite{bib:4596} are widely used not only for analyses of chemical bonds but also for accurate calculations of anomalous Hall conductivity and transport properties.

The generic expression of a WO is
\begin{gather}
    w_{\boldsymbol{R} n} (\boldsymbol{r})
    =
        \frac{1}{N_k}
        \sum_{\boldsymbol{k}, p}
        e^{-i \boldsymbol{k} \cdot \boldsymbol{R}}
        \psi_{\boldsymbol{k} p} (\boldsymbol{r})
        U_{p n}^{(\boldsymbol{k})}
    .
    \label{def_Wannier}
\end{gather}
$\boldsymbol{R}$ is the lattice point where the unit cell containing the $n$th WO is located.
$U^{(\boldsymbol{k})}$ is a unitary matrix at $\boldsymbol{k}$ for the construction of localized orbitals from the extending Bloch orbitals $\psi_{\boldsymbol{k} p} (\boldsymbol{r})$.
When the transformation matrix $U^{(\boldsymbol{k})}$ is identity at each $\boldsymbol{k}$,
the normal WOs (NWOs)\cite{bib:5} are obtained.
When the matrices are determined so that the spread functional\cite{bib:369, bib:368} is minimized,
on the other hand,
the MLWOs are obtained.

\subsubsection{Direct interpolation}

The Bloch sum of the localized orbital in eq. (\ref{def_Wannier}) for a wave vector $\boldsymbol{k}$ is defined as
$
    w_{\boldsymbol{k} n} (\boldsymbol{r})
    =
        \sum_{\boldsymbol{R}}
            e^{i \boldsymbol{k} \cdot \boldsymbol{R}}
            w_{\boldsymbol{R} n} (\boldsymbol{r})
    ,
$
which extends over the whole crystal.
The Bloch sums of the target bands allows one to transform the CCSD GF in the band representation,
which is also said to be in the Bloch gauge,
to the new one in the Wannier gauge as
\begin{gather}
   	G_{n n'} (\boldsymbol{k}, \omega)
    =
    	\sum_{p, p'}
            (U^{(\boldsymbol{k}) \dagger})_{n p}
			G_{p p'} (\boldsymbol{k}, \omega)
            U_{p' n'}^{(\boldsymbol{k})}
    .
	\label{GF_wan_from_band_repr}
\end{gather}
For the calculated GF at $N_k$ sampled $k$ points in the Brillouin zone (BZ),
we perform Fourier transformation as 
\begin{gather}
   	\widetilde{G}_{n n'} (\boldsymbol{R}, \omega)
    =
        \frac{1}{N_k}
    	\sum_{\boldsymbol{k}}^{\mathrm{sampled}}
            e^{-i \boldsymbol{k} \cdot \boldsymbol{R}}
           	G_{n n'} (\boldsymbol{k}, \omega)
    ,
    \label{def_G_R_sampled}
\end{gather}
which is ideally equal to the exact Fourier transform $G_{n n'} (\boldsymbol{R}, \omega)$ in the limit of an infinite number of sampled $k$ points.
The real-space representation defined above enables us to obtain the GF for an arbitrary wave vector via inverse Fourier transformation as
\begin{gather}
   	\widetilde{G}_{n n'}^{\mathrm{d}} (\boldsymbol{k}, \omega)
   	=
    	\sum_{\boldsymbol{R}}
        	e^{i \boldsymbol{k} \cdot \boldsymbol{R} }
		   	\widetilde{G}_{n n'} (\boldsymbol{R}, \omega)
    ,
    \label{def_G_R_direct}
\end{gather}
which we call the direct interpolation hereafter.

It is clear from eq. (\ref{def_spec_k}) that
the spectral function $\widetilde{A}^{\mathrm{d}} (\boldsymbol{k}, \omega)$ calculated from direct interpolation does not depend on the matrices $U^{(\boldsymbol{k})}$ since they are unitary.
It is also clear from eq. (\ref{def_G_R_direct}) that
the interpolated spectral function integrated over an arbitrarily fine $k$ mesh is identical to the original spectra integrated over the sampled $k$ points:
$\widetilde{A}^{\mathrm{d}} (\omega) = A (\omega)$.

\subsubsection{Self-energy-mediated interpolation}

We cannot avoid being concerned about the reliability of $\widetilde{G}_{n n'} (\boldsymbol{R}, \omega)$ defined in eq. (\ref{def_G_R_sampled}) since
the number of sampled $k$ points has to be small in general due to 
the large computational cost of CCSD and subsequent GF calculations.
To circumvent the difficulty in increasing the number of sampled $k$ points,
we propose another interpolation scheme for GFs here.

The self-energy $\Sigma$ is obtained via the Dyson equation
\begin{gather}
    G^{-1} (\boldsymbol{k}, \omega)
    =
        G_0^{-1} (\boldsymbol{k}, \omega)
        -
        \Sigma (\boldsymbol{k}, \omega)
    ,
    \label{Dyson_eq}
\end{gather}
where $G_0$ is the HF GF.
Substituting the CCSD GF in eq. (\ref{def_CCSD_GF}) into the matrix equation above,
we get the CCSD self-energy.
It is noted here that the CCSD self-energy does not contain the contributions from the HF self-energy diagrams,
which are already contained in $G_0$.\cite{stefanucci2013nonequilibrium}
The HF GF in the Bloch gauge is diagonal in reciprocal space,
whose component is given by
\begin{gather}
    (G_0^{-1})_{p p'} (\boldsymbol{k}, \omega)
    =
        ( \omega - \varepsilon_{\boldsymbol{k} p} )
        \delta_{p p'}
        ,
\end{gather}
where $\varepsilon_{\boldsymbol{k} p}$ is the HF orbital energy.

The interpolation procedure is as follows.
We first calculate the CCSD self-energy in the Bloch gauge via eq. (\ref{Dyson_eq}),
which is then transformed into the Wannier gauge as well as in eq. (\ref{GF_wan_from_band_repr}).
We apply Fourier transformation to it using the sampled $k$ points to get $\widetilde{\Sigma}_{n n'} (\boldsymbol{R}, \omega)$ similarly to eq. (\ref{def_G_R_sampled}).
From this real-space representation,
we can interpolate the self-energy $\widetilde{\Sigma}_{n n'} (\boldsymbol{k}, \omega)$ for an arbitrary wave vector via inverse Fourier transformation,
which we plug into the Dyson equation to get the interpolated GF
\begin{gather}
   	\widetilde{G}^{\mathrm{sem}} (\boldsymbol{k}, \omega)
    =
    	[
	   	\widetilde{G}_0^{-1} (\boldsymbol{k}, \omega)
        -
	   	\widetilde{\Sigma} (\boldsymbol{k}, \omega)
        ]^{-1}
    .
\end{gather}
We call this scheme the self-energy-mediated interpolation hereafter.
Since this scheme includes inversion of matrices,
the resultant spectral function depends on the construction of WOs since the unitary matrices $U^{(\boldsymbol{k})}$ depend on $\boldsymbol{k}$ in general.

There exists an attempt for interpolating $GW$ quasiparticle band structure using MLWOs done by Hamann and Vanderbilt.\cite{bib:4684}
Their scheme uses the $GW$ quasiparticle wave functions and their orbital energies to get the $GW$ Hamiltonian in real space by adopting a manner computationally similar to our direct interpolation.
Their formalism for efficient interpolation of correlated band structure stems from the localized shapes of MLWOs.
The self-energy-mediated interpolation, on the other hand,
relies on the localized nature of self-energies,
as will be demonstrated later.
It will be interesting to examine the interpolation using the $GW$ self-energy in the future.

\section{computational details}
\label{sec:comp_details}

We adopt STO-3G basis set for the Cartesian Gaussian-type basis functions\cite{Helgaker} of all the elements in the present study.
The Coulomb integrals between AOs are calculated efficiently.\cite{Libint1}
By transforming them using the results of the HF calculations for periodic systems,
we obtain the integrals between the Bloch orbitals\cite{bib:4024},
with which we perform the CCSD calculations by successive substitution.
We solve the IP-EOM-CCSD and EA-EOM-CCSD equations in eqs. (\ref{def_ip_eom}) and (\ref{def_ea_eom}), respectively,
by using the shifted BiCG method.\cite{bib:4514,bib:4515,bib:4512}
We set $\delta = 0.02$ Ht in eq. (\ref{def_spec_k}) throughout this study.
For the construction of MLWOs,
we calculate the overlaps between the cell-periodic parts of the Bloch orbitals as input to wannier90.\cite{bib:4587}

\section{results and discussion}
\label{sec:results}

\subsection{LiH chain}

\subsubsection{Band structure and CCSD GF}

For a LiH chain composed of equidistant atoms,
we first optimized the lattice constant via HF calculations using $N_k = 12 \times 1 \times 1$ sampled $k$ points.
We obtained the optimized lattice constant $a = 3.28$ \AA,
in reasonable agreement with previous studies.\cite{bib:4586,bib:4585}
We obtained a restricted HF (RHF) solution for this lattice constant and
adopted it as the reference state for the CCSD calculation.

\begin{figure}
\begin{center}
\includegraphics[width=7cm]{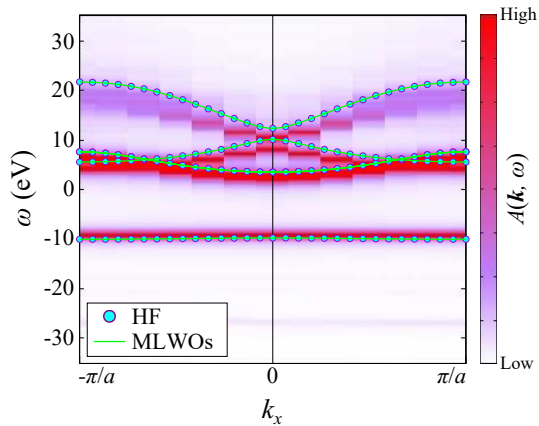}
\end{center}
\caption{
HF band structure of a LiH chain as circles and that obtained with the MLWOs as curves.
The spectral function $A (\boldsymbol{k}, \omega)$ calculated from the CCSD GF at 12 sampled $k$ points are also shown. 
The chain extends in the $x$ direction.
}
\label{Fig_LiH_bands}
\end{figure}

We constructed the MLWOs from all the 6 bands.
The MLWOs can be used for interpolation of the original bands.\cite{bib:369, bib:368}
The HF bands and their Wannier interpolation are plotted in Fig. \ref{Fig_LiH_bands},
where the original bands are accurately reproduced.
The flat valence band at $\omega = -10$ eV comes from the H 1$s$ orbital,
while the conduction bands are dispersive.
The CCSD spectral function $A (\boldsymbol{k}, \omega)$ is also shown in the figure.
We find clear correspondence between the HF band energies and the quasiparticle peaks in the CCSD spectra.
In addition, low intensities exist in the CCSD spectra,
known as the satellite peaks.\cite{bib:4473}
They are direct consequences of many-body effects taken into account by the correlated approach.
The locations of quasiparticle peaks below (above) the Fermi level are closer to $\omega = 0$ than those of the valence (conduction) HF band energies are,
as generic characteristics of correlation effects.
Since the system is spin unpolarized,
the spectral intensities are the same at an arbitrary $\boldsymbol{k}$ and $-\boldsymbol{k}$ due to time reversal symmetry.

\subsubsection{Direct interpolation}

\begin{figure}
\begin{center}
\includegraphics[width=7cm]{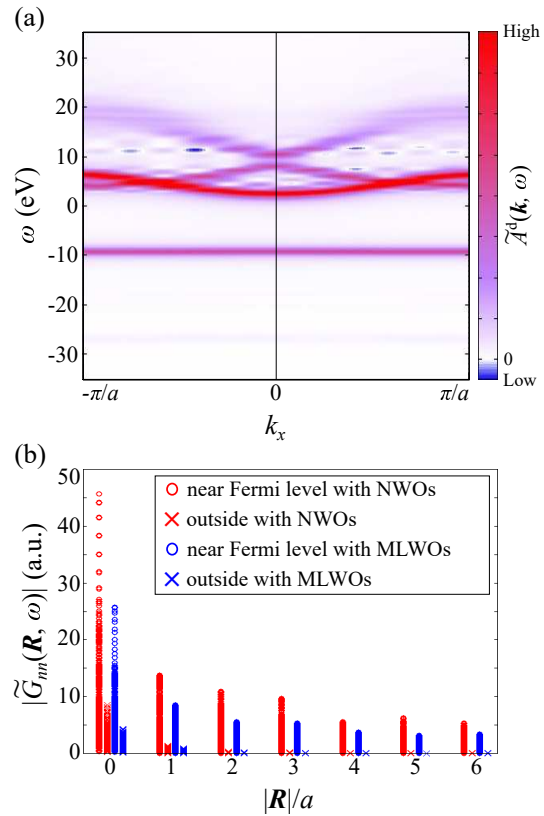}
\end{center}
\caption{
(a)
Spectral function $\widetilde{A}^{\mathrm{d}} (\boldsymbol{k}, \omega)$ calculated from the direct interpolation of CCSD GF for a LiH chain.
(b) 
The absolute values $|\widetilde{G}_{n n} (\boldsymbol{R}, \omega)|$ of diagonal components of the GFs as functions of $|\boldsymbol{R}|$.
Those obtained using the NWOs and MLWOs for the energy region near the Fermi level ($-12$ eV $< \omega < $ 22 eV) and the outside region are plotted.
}
\label{Fig_LiH_gf_direct}
\end{figure}

The spectral function $\widetilde{A}^{\mathrm{d}} (\boldsymbol{k}, \omega)$ calculated from direct interpolation is shown in Fig. \ref{Fig_LiH_gf_direct} (a).
One finds soon that three obviously unfavorable features exist in the interpolated spectra.
First,
the quasiparticle peaks for the highest conduction band consist of spots separated by the distance $\Delta k_x$ between the neighboring sampled $k$ points.
Second,
there exist trains of specks at $\omega = 10$ and 5 eV, where each speck is separated by $\Delta k_x$ again.
The spectral intensities for some of the specks are,
even worse, unphysically negative. 
Third, 
the time reversal symmetry is not preserved in the spectra,
particularly for the trains of specks.

For the sampled frequencies in a range $-40$ eV $< \omega < 40$ eV,
the absolute values of diagonal components of
$\widetilde{G} (\boldsymbol{R}, \omega)$
in the region near the Fermi level
($-12$ eV $< \omega < 22$ eV) and
the outside region are plotted in Fig. \ref{Fig_LiH_gf_direct} (b).
Although the decreasing tendencies of those values for the frequencies near the Fermi level are seen for both kinds of WOs,
their convergence is slow for the increase in $|\boldsymbol{R}|$.
In contrast,
the diagonal components for the other frequencies decrease rapidly enough already at $|\boldsymbol{R}|/a = 2$.
These observations indicate that the sampled $k$ points are too few for the direct interpolation near the Fermi level despite the fact that the HF bands are sufficiently convergent with respect to the $k$ points.
The unfavorable features of the direct-interpolated spectra enumerated above are numerical artifacts due to the insufficient number of sampled $k$ points.

\subsubsection{Self-energy-mediated interpolation}

\begin{figure}
\begin{center}
\includegraphics[width=7cm]{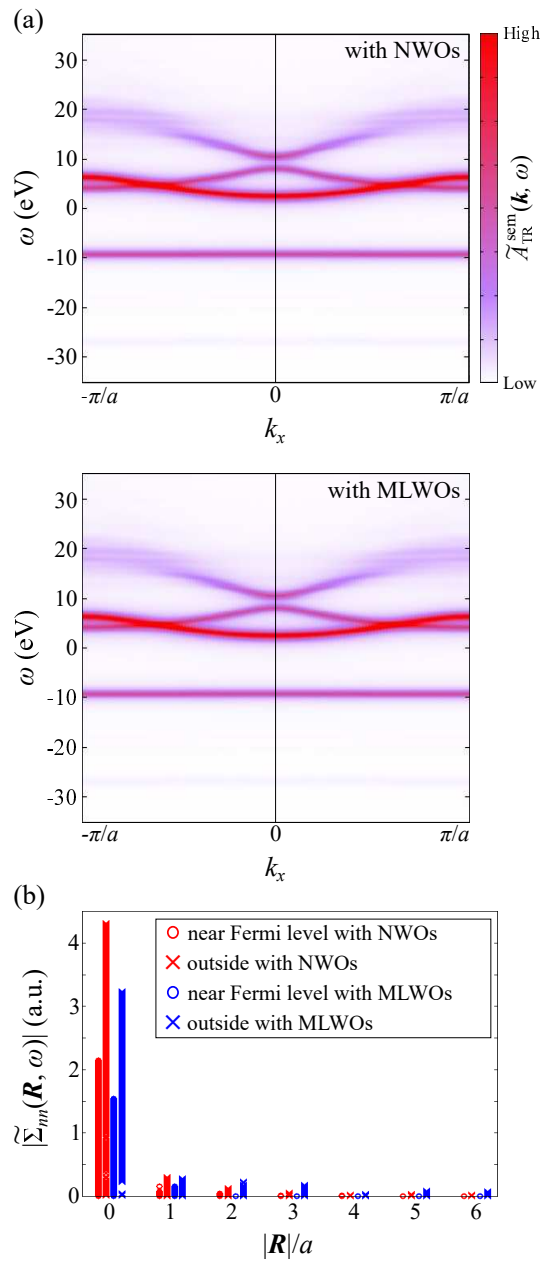}
\end{center}
\caption{
(a)
Spectral functions $\widetilde{A}_{\mathrm{TR}}^{\mathrm{sem}} (\boldsymbol{k}, \omega)$ calculated from the self-energy-mediated interpolation for a LiH chain by using the NWOs and MLWOs are shown in the upper and lower panels, respectively.
(b) 
The absolute values $|\widetilde{\Sigma}_{n n} (\boldsymbol{R}, \omega)|$ of diagonal components of the self-energies as functions of $|\boldsymbol{R}|$.
}
\label{Fig_LiH_gf_dyson}
\end{figure}

To circumvent the direct interpolation,
let us next try the self-energy-mediated interpolation.
We impose the time reversal symmetry condition on the spectral function from the self-energy-mediated interpolation as
\begin{gather}
   	\widetilde{A}^{\mathrm{sem}}_{\mathrm{TR}} (\boldsymbol{k}, \omega)
   	\equiv
   	    \frac{
   	        \widetilde{A}^{\mathrm{sem}} (\boldsymbol{k}, \omega)
       	    +
           	\widetilde{A}^{\mathrm{sem}} (-\boldsymbol{k}, \omega)
           	}{2}
    .
    \label{def_time_rev_spec}
\end{gather}
The spectral functions calculated in this way
by using the NWOs and MLWOs are shown in Fig. \ref{Fig_LiH_gf_dyson} (a),
where the unfavorable features for the direct interpolation do not appear.
The spectra for the two kinds of NWOs are almost indistinguishable from each other.
The absolute values of diagonal components of
$\widetilde{\Sigma} (\boldsymbol{R}, \omega)$
in the same regions as in Fig. \ref{Fig_LiH_gf_direct} (b) are plotted in Fig. \ref{Fig_LiH_gf_dyson} (b).
Those values decrease rapidly enough already at $|\boldsymbol{R}|/a = 1$ for all the frequencies.
This means that the number of sampled $k$ points is sufficient for the description of the variation in CCSD self-energy in reciprocal space,
and hence the self-energy-mediated interpolation of GF is reliable within the accuracy ensured by our preceding procedure of CCSD GF calculations.

\begin{figure}
\begin{center}
\includegraphics[width=7cm]{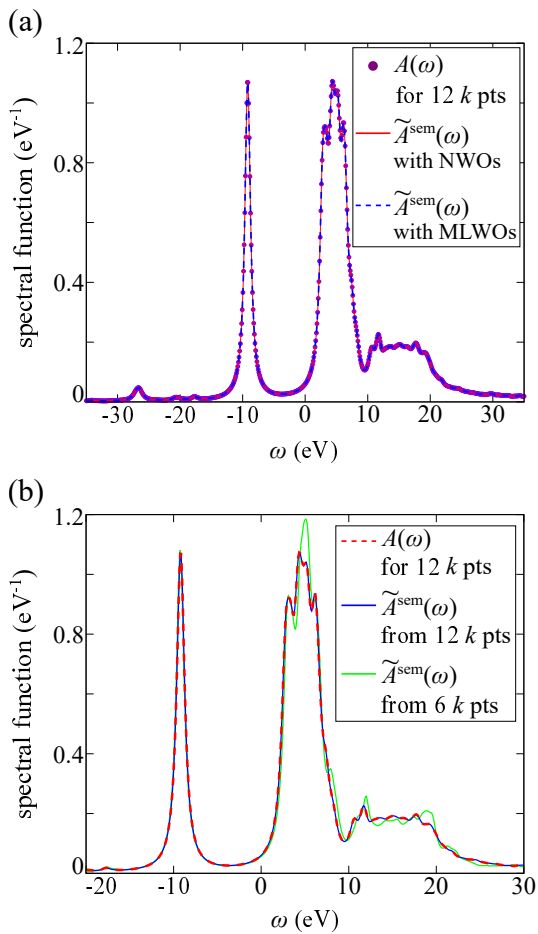}
\end{center}
\caption{
(a)
$k$-integrated spectral functions of a LiH chain for
the original CCSD GF at 12 sampled $k$ points and
the interpolated GFs using the WOs.
(b)
The original spectra and the self-energy-mediated interpolated ones using the NWOs for 12 sampled $k$ points.
The latter for 6 sampled $k$ points are also shown.
}
\label{Fig_LiH_spec_integ}
\end{figure}

The spectral functions integrated over $k$ points,
or equivalently the densities of states,
for the original CCSD GF and
the interpolated GFs using the WOs are shown in Fig. \ref{Fig_LiH_spec_integ} (a).
Those for the two kinds of WOs look indistinguishable,
in addition to which they almost coincide with the original spectra.

To see whether the self-energy-mediated interpolation using a smaller number of sampled $k$ points reproduces the original spectra,
we calculated the interpolated spectra for $N_k = 6$ and plotted them in Fig. \ref{Fig_LiH_spec_integ} (b).
The interpolated spectra from $N_k = 12$ and those from $N_k = 6$ look quite similar to each other,
implying the usefulness of our scheme for $k$-integrated spectra.

\subsection{$trans$-polyacetylene}

\subsubsection{Band structure and CCSD GF}

For $trans$-polyacetylene,
we adopted the structural parameters provided by Teramae\cite{bib:4040} to construct the unit cell consisting of two C atoms and two H atoms,
where the bond alternation has occurred.\cite{bib:4566,bib:4565}
We obtained an RHF solution for this geometry using $N_k = 8 \times 1 \times 1$ sampled $k$ points and
adopted it as the reference state for the CCSD calculations.
Although it has been shown\cite{bib:4602} that
the band picture on this system is dubious by resorting to DFT calculations incorporating the zero-point vibrations of atoms,
we keep to the band picture since the main purpose of present study is to propose the interpolation schemes.

\begin{figure}
\begin{center}
\includegraphics[width=7cm]{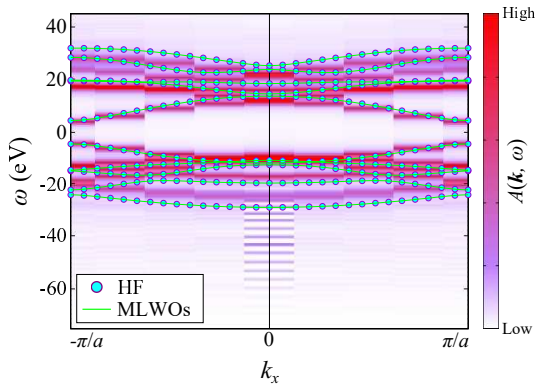}
\end{center}
\caption{
HF band structure of $trans$-polyacetylene as circles and that obtained with the MLWOs as curves.
The spectral function $A (\boldsymbol{k}, \omega)$ calculated from the CCSD GF at 8 sampled $k$ points are also shown. 
The polymer extends in the $x$ direction.
$a$ is the lattice constant.
}
\label{Fig_C2H2_bands}
\end{figure}

We constructed the MLWOs from the 10 bands near the Fermi level.
The HF bands and their Wannier interpolation are plotted in Fig. \ref{Fig_C2H2_bands},
where the original bands are accurately reproduced.
The calculated band gap of 8.9 eV at X ($k_x = \pm \pi/a$) is in reasonable agreement obtained by Teramae\cite{bib:550} using the same basis set.
These calculated gaps are much larger than the experimental ones\cite{bib:4565,bib:4564} of 1 - 2 eV,
as is often the case with HF calculations.
The CCSD spectral function is also shown in the figure,
where the satellite peaks for $\Gamma$ ($k_x = 0$) have stronger intensities than for $k_x \ne 0$.

\subsubsection{Direct interpolation}

\begin{figure}
\begin{center}
\includegraphics[width=7cm]{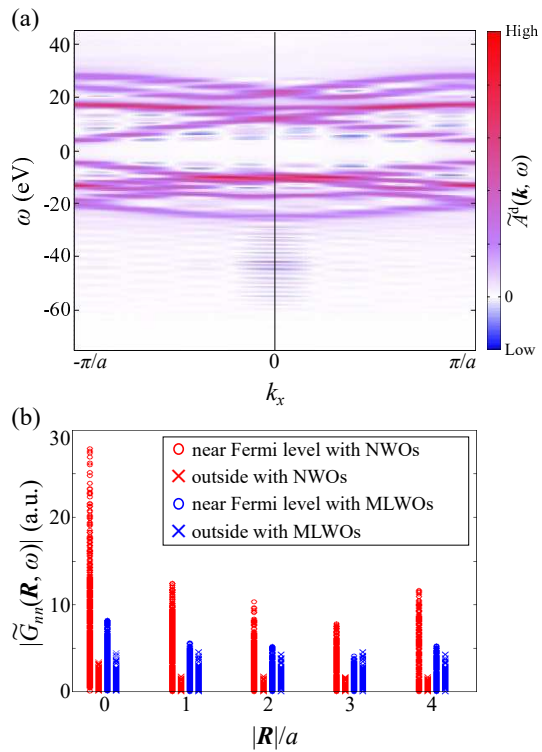}
\end{center}
\caption{
(a)
Spectral function $\widetilde{A}^{\mathrm{d}} (\boldsymbol{k}, \omega)$ calculated from the direct interpolation of CCSD GF for $trans$-polyacetylene.
(b)
The absolute values $|\widetilde{G}_{n n} (\boldsymbol{R}, \omega)|$ of diagonal components of the GFs as functions of $|\boldsymbol{R}|$.
Those obtained using the NWOs and MLWOs for the energy region near the Fermi level ($-33$ eV $< \omega < $ $33$ eV) and the outside region are plotted.
}
\label{Fig_C2H2_gf_direct}
\end{figure}

The spectral function $\widetilde{A}^{\mathrm{d}} (\boldsymbol{k}, \omega)$ calculated from direct interpolation is shown in Fig. \ref{Fig_C2H2_gf_direct} (a),
where one finds unfavorable features similarly to the case of a LiH chain.
For the sampled frequencies in a range $-60$ eV $< \omega < 50$ eV,
the absolute values of diagonal components of
$\widetilde{G} (\boldsymbol{R}, \omega)$
in the region near the Fermi level
($-33$ eV $< \omega < $ $33$ eV) and
the outside region are plotted in Fig. \ref{Fig_C2H2_gf_direct} (b).
No clear tendency of decrease in those values is seen for the two kinds of WOs.
The numerical artifacts in the direct-interpolated spectra thus look more prominent than for a LiH chain.
In particular,
the interpolated satellite peaks for $\omega < -25$ eV can be unphysically negative,
as seen in Fig. \ref{Fig_C2H2_gf_direct} (a).

\subsubsection{Self-energy-mediated interpolation}

\begin{figure}
\begin{center}
\includegraphics[width=7cm]{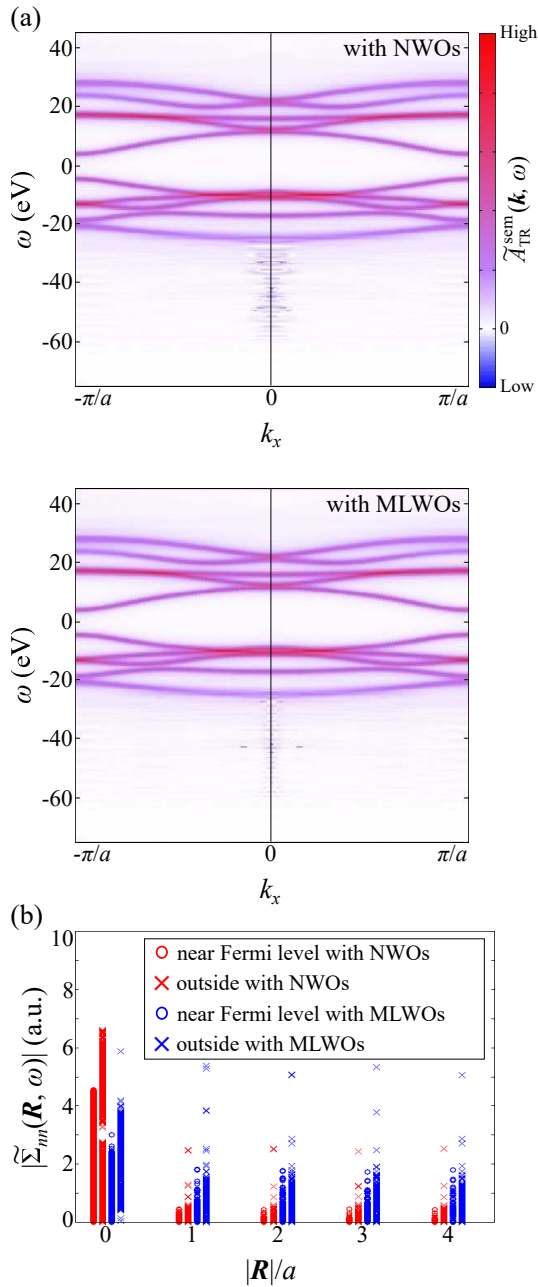}
\end{center}
\caption{
(a)
Spectral functions $\widetilde{A}_{\mathrm{TR}}^{\mathrm{sem}} (\boldsymbol{k}, \omega)$ calculated from the self-energy-mediated interpolation for $trans$-polyacetylene by using the NWOs and MLWOs are shown in the upper and lower panels, respectively.
(b) 
The absolute values $|\widetilde{\Sigma}_{n n} (\boldsymbol{R}, \omega)|$ of diagonal components of the self-energies as functions of $|\boldsymbol{R}|$.
}
\label{Fig_C2H2_gf_dyson}
\end{figure}

The spectral functions
$\widetilde{A}^{\mathrm{sem}}_{\mathrm{TR}} (\boldsymbol{k}, \omega)$
calculated via self-energy-mediated interpolation by using the NWOs and MLWOs are shown in Fig. \ref{Fig_C2H2_gf_dyson} (a).
Unphysical intensity does not appear in the interpolated spectra near the Fermi level.
The absolute values of diagonal components of
$\widetilde{\Sigma} (\boldsymbol{R}, \omega)$
in the same frequency regions as in Fig. \ref{Fig_C2H2_gf_direct} (b) are plotted in Fig. \ref{Fig_C2H2_gf_dyson} (b).
The diagonal components near the Fermi level for the NWOs are large for $|\boldsymbol{R}| = 0$ compared to $|\boldsymbol{R}| \ne 0$.
This is also the case for the MLWOs.
On the other hand,
there exist significant contributions from
$|\boldsymbol{R}| \ne 0$ for the frequencies far from the Fermi level in contrast to the case of a LiH chain.
The unphysical intensities are thus seen for $-25$ eV $< \omega <$ $60$ eV at $\Gamma$,
where the two kinds of WOs give slightly different spectra. [See Fig. \ref{Fig_C2H2_gf_dyson} (a)]

\begin{figure}
\begin{center}
\includegraphics[width=7cm]{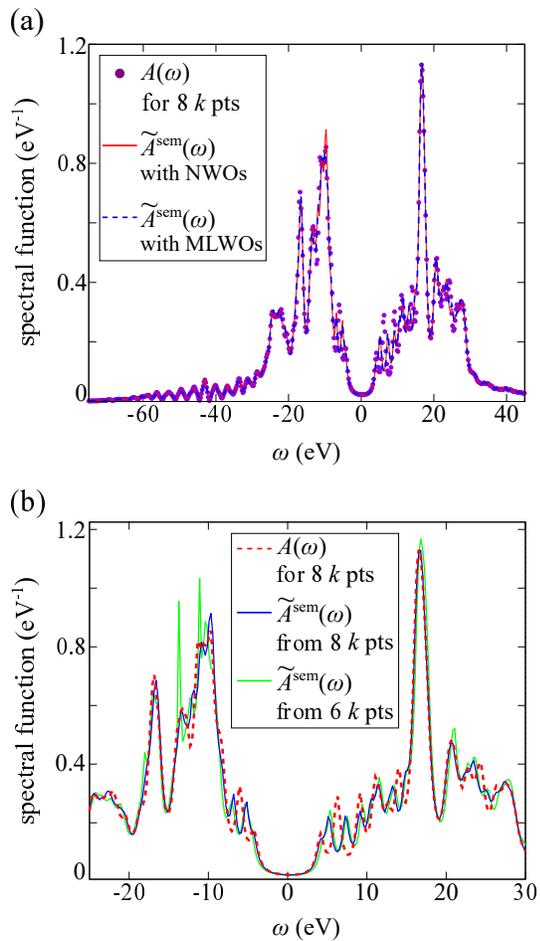}
\end{center}
\caption{
(a)
$k$-integrated spectral functions of $trans$-polyacetylene for
the original CCSD GF at 8 sampled $k$ points and
the interpolated GFs using the WOs.
(b)
The original spectra and the self-energy-mediated interpolated ones using the NWOs for 8 sampled $k$ points.
The latter for 6 sampled $k$ points are also shown.
}
\label{Fig_C2H2_spec_integ}
\end{figure}

The spectral functions integrated over $k$ points for the original CCSD GF and the interpolated GFs using the WOs are shown in Fig. \ref{Fig_C2H2_spec_integ} (a).
Those for the two kinds of WOs look indistinguishable even for $\omega < -25$ eV in contrast to the $k$-resolved spectra. [See Fig. \ref{Fig_C2H2_gf_dyson} (a)]
Furthermore, negative intensities do not appear for those frequencies in the $k$-integrated spectra.
These observations imply that accurate interpolation of $k$-resolved spectra requires more sampled $k$ points than $k$-integrated spectra do.

To see whether the self-energy-mediated interpolation using a small number of sampled $k$ points allows one to access the $k$-integrated spectra which would be obtained for a larger number of $k$ points,
we calculated the interpolated spectra for $N_k = 6$ and plotted them in Fig. \ref{Fig_C2H2_spec_integ} (b).
The interpolated spectra from $N_k = 8$ and those from $N_k = 6$ look quite similar to each other,
indicative of well converged self-energy with respect to $N_k$.
On the other hand,
the peak locations of the original spectra for $-10$ eV $< \omega < 15$ eV differ slightly from those of the interpolated spectra,
implying slow convergence of the original GF.
These results corroborate the usefulness of the self-energy-mediated interpolation scheme as well as in the LiH chain case.

It has been demonstrated that the self-energy-mediated interpolation is successful for our two systems at least near the Fermi level.
Our results are consistent with the often adopted assumption that the self-energy of an electronic system is more localized than the GF.
The dynamical mean-field theory (DMFT)\cite{PhysRevB.45.6479} and its application in electronic-structure calculations\cite{bib:3127} are based on this assumption and have been used successfully.

\section{conclusions}
\label{sec:conclusions}

We proposed two schemes for interpolation of the one-particle GF calculated within CCSD method for a periodic system.
These schemes employ transformation of representation from reciprocal to real spaces by using WOs for circumventing huge cost for a large number of sampled $k$ points.
One of the schemes is the direct interpolation,
which obtains the GF straightforwardly by using Fourier transformation.
The other is the self-energy-mediated interpolation,
which obtains the GF via the Dyson equation.
We applied the schemes to two insulating systems,
a LiH chain and $trans$-polyacetylene,
and examined their validity in detail.
We found that the direct-interpolated GFs suffered from numerical artifacts stemming from slow convergence of CCSD GFs in real space.
The self-energy-mediated interpolation, on the other hand,
was found to provide more physically appropriate GFs
due to the localized nature of CCSD self-energies.
We should keep in mind that in a metallic system,
whose density matrix\cite{bib:4597,bib:4598} and GF\cite{bib:4604} decay only algebraically at a zero temperature,
a large number of sampled $k$ points would be required for sufficiently convergent results.
Remembering the widely accepted assumption that the self-energy of an interacting system is more localized than the GF,
the self-energy-mediated interpolation is expected to be more suitable for generic systems than the direct interpolation.

Since our interpolation schemes are not restricted to CCSD method,
they are applicable to any correlated methods in quantum chemistry as long as it provides a way to obtain one-particle GFs.
Development of various correlated methods with GFs in solids is thus important for reliable explanations and predictions of their spectral shapes and excitation energies.

\begin{acknowledgments}

This research was supported by MEXT as Exploratory Challenge on Post-K computer (Frontiers of Basic Science: Challenging the Limits).
This research used computational resources of the K computer provided by the RIKEN Advanced Institute for Computational Science through the HPCI System Research project (Project ID: hp180227).

\end{acknowledgments}

\bibliographystyle{apsrev4-1}
\bibliography{ref}

\end{document}